\documentclass[12pt]{article}
\usepackage{amssymb}
\usepackage{amsmath}

\oddsidemargin  - 5 mm
\evensidemargin - 5 mm

\input{tcilatex}

\begin{document}

\date{}
\title{Block-Structure Method for the Solution of the Matrix System of
Equations $g_{ij}g^{jk}=\delta _{i}^{k}$ in the $N-$dimensional Case }
\author{Bogdan G. Dimitrov \thanks{%
Electronic mail: bogdan@theor.jinr.ru} \\
Bogoliubov Laboratory for Theoretical Physics\\
Joint Institute for Nuclear Research \\
6 Joliot-Curie str. \\
Dubna 141980, Russia}
\maketitle

\begin{abstract}
\ \ 

\ \ \ In this paper a new block-structure method is presented for the
solution of the well-known from gravity theory matrix system of equations $%
g_{ij}g^{jk}=\delta _{i}^{k}$ (with respect to the unknown covariant
components $g_{ij}$ and by known contravariant ones $g^{jk}$) by
transforming this matrix system into a linear algebraic system of equations
in the general $N-$dimensional case. Although powerful computer methods
exist for the solution of this problem for a given (fixed) dimension of the
matrices $g^{ij}$ and especially for numerical elements of $g^{ij}$, the
structure of the obtained linear algebraic system in the general $N-$%
dimensional case and for arbitrary elements of $g^{ij}$ (functions) has not
been known.

The proposed new \ analytical \ block-structure method for the case of
symmetrical matrices $g_{ij}$ and $g^{jk}$ (the standard case in gravity
theory) is based on the construction of a block-structure matrix, whose
"elements" are again matrices. The method allows to obtain the structure of
this linear system in the general $N-$dimensional case, after multiplication
(to the left) with the transponed matrix. \ 

Some arguments are given why \ the proposed method may be applied, after
some refinement and generalization for the case of non-symmetrical matrices $%
g_{ij}$ and $g^{jk}$, for finding the graviton modes in the Kaluza-Klein
expansion in theories with extra dimensions.
\end{abstract}

\bigskip

\vskip .5cm

\section{\protect\bigskip INTRODUCTION}

The system of equations $g_{ij}g^{jk}=\delta _{i}^{k}$ is well-known and
important in gravity theory, since from it the contravariant metric tensor
components are determined under known covariant ones. In spite of its
importance, this system is not well-understood as a mathematical object. For
example, in Bergman's book on General Relativity [1] (Ch. 5, eq. 5.64) it is
written that \textit{" if the determinant }$g_{ij}$\textit{\ is not equal to
zero, then a multitude of new variables }$g^{jk\text{ }}$\textit{can be
introduced according to the relation }$g_{ij}\widetilde{g}^{jk}=\delta
_{i}^{k}$\textit{"}. No doubt, similar statements can be seen in other
books. Bergmann's statement would have been true if this system is a trivial
linear algebraic system of equations with a number of equations equal to the
number of variables. But in fact, this is a \textit{predetermined system of
equations} with a number of equations $n^{2}$ greater than the number $%
\left( 
\begin{array}{c}
n \\ 
2%
\end{array}%
\right) +n$ of unknown variables. This is the first reason why Bergmann' s
statement does not hold. Secondly, \textit{this is not a linear algebraic
system of equations, but an operator (matrix) system of equations} [2] of
the type $AX=B$, where $X$ is no longer a vector-column, but a matrix.
Consequently, the linear algebra theorems do not hold any more.

One of the\ basic\ new\ results\ in\ the\ present\ paper\ is\ that\ the\
initially\ given\ operator\ system\ of\ equations\ can\ be\ reduced\ to\ a\
linear\ algebraic\ system\ of\ equations\ $B\widetilde{Y}=T$, where the
unknown elements $\widetilde{Y}$ represent the matrix elements $X$, arranged
in the form of a vector-column. At first, the system will be written in the
form $\widetilde{A}\widetilde{Y}=\delta _{i}^{k}$, where $\widetilde{A}$ is
a non-quadratic $\left[ \left( 
\begin{array}{c}
n \\ 
2%
\end{array}%
\right) +n\right] \times n^{2}$ matrix. Further, non-quadratic matrices will
be denoted as $a\times b$, where $a$ will be the number of rows in the
matrix and $b$ will be the number of columns.

In the conclusion, a possible application of the proposed mathematical
method will be briefly commented in reference to the problem about finding
the graviton's modes in the Kaluza-Klein's expansion under toroidal
compactification in theories with extra dimensions.

\section{ Some general properties of the system $g_{ij}\widetilde{g}^{jk}=%
\protect\delta _{i}^{k}$ in the general $n-$dimensional case.}

Since the number of equations is greater than the number of variables, from
the $n^{2}$ equations one can select $n^{2}-n$ equations with different
values of $\ i$ and $k$, for which the right-hand side (R. H. S.) will be
zero. Yet the number of the chosen equations remains to be greater than the
number of variables $\left( 
\begin{array}{c}
n \\ 
2%
\end{array}%
\right) +n$, which is confirmed by the equality 
\begin{equation}
(n^{2}-n)-\left[ \left( 
\begin{array}{c}
n \\ 
2%
\end{array}%
\right) +n\right] =\frac{n(n-3)}{2}\text{ \ \ \ ,}  \tag{2.1}
\end{equation}%
fulfilled for $n>3$. Only for the case $n=3$ , the number of the equations $%
n^{2}-n=6$ with a zero R. H. S. becomes exactly equal to the number of
variables $\left( 
\begin{array}{c}
n \\ 
2%
\end{array}%
\right) +n=\frac{n(n+1)}{2}=\frac{3.\text{ }4}{2}=6$. Therefore for the case
of an arbitrary $n$ from these $n^{2}-n$ equations one can choose $\left[
\left( 
\begin{array}{c}
n \\ 
2%
\end{array}%
\right) +n\right] $ equations. In each row of the $\left[ \left( 
\begin{array}{c}
n \\ 
2%
\end{array}%
\right) +n\right] \times \left[ \left( 
\begin{array}{c}
n \\ 
2%
\end{array}%
\right) +n\right] $ matrix of coefficient functions of this system there
will be only $n$ functions $\widetilde{g}^{jk}$, the rest $\left( 
\begin{array}{c}
n \\ 
2%
\end{array}%
\right) $ of the \ elements will be zero. The determinant of the matrix will
be equal to zero, a proof of which in the general case of an arbitrary $n$
(and also for $n=3$) will be presented in the next sections. Therefore, the
solutions $g_{ij}$ of this $\left( 
\begin{array}{c}
n \\ 
2%
\end{array}%
\right) +n$ dimensional homogeneous \ system of equations with a zero
determinant are arbitrary.

Now we are left with $\frac{n(n-3)}{2}$ equations (again with \ $\left( 
\begin{array}{c}
n \\ 
2%
\end{array}%
\right) +n$ variables) with a zero R. H. S. (we shall call it the first
system of equations) and with another $n$ equations (the second system)\
with a R. H. S. of each equation, equal to $1$. Since the number of
variables $\left( 
\begin{array}{c}
n \\ 
2%
\end{array}%
\right) +n$ in the first system is greater than the number $\frac{n(n-3)}{2}$
of equations and 
\begin{equation}
\left( 
\begin{array}{c}
n \\ 
2%
\end{array}%
\right) +n-\frac{n(n-3)}{2}=2n\text{ \ \ ,}  \tag{2.2}
\end{equation}%
again one can treat as unknown only $\frac{n(n-3)}{2}$ variables and
transfer the rest $2n$ variables in the R. H. S., which will become
different from zero. Analogously, for the second system one can treat as
unknown only $n$ variables and transfer the rest $\left( 
\begin{array}{c}
n \\ 
2%
\end{array}%
\right) =$ $\frac{n(n-1)}{2}$ variables in the R. H. S. If this R. H. S. is
different from zero and moreover, the determinant of the coefficient
functions $\widetilde{g}^{ij}$ is also different from zero, then one can
find unique solutions for these $n$\ variables $g_{ij}$. Now comes the most
important point of the proof: Since 
\begin{equation}
n+\frac{n(n-1)}{2}>2n\text{ \ \ \ \ ,}  \tag{2.3}
\end{equation}%
one can take all $n$ of the \ uniquely found solutions plus $n$ more unfixed
(freely varied) variables from the R. H.S. of the second system and
\textquotedblright plunge\textquotedblright\ them into the R. H. S. of the
first system. Let us remember that the first system has $\frac{n(n-3)}{2}$
unknown variables, but one may note that $\left( 
\begin{array}{c}
n \\ 
2%
\end{array}%
\right) +n-2n=\frac{n(n-3)}{2}$ variables from the second system have not
been transferred in the R. H. S. of the first system. Therefore, one can
choose these $\frac{n(n-3)}{2}$ variables to be the unknown variables for
the first system, and if the determinant of the coefficient functions is
nonzero, an unique solution can be found for them.

As a whole , one would have a maximum of $n+$\ $\frac{n(n-3)}{2}=\frac{n(n-1)%
}{2}$\ uniquely fixed variables and the rest of the variables may be varied
freely.

\section{\ BLOCK STRUCTURE METHOD FOR \ THE \ PARTIAL $n=3$ \ CASE}

\bigskip The purpose of the present section will be to develop a method for
solving the system of equations $g_{ij}g^{jk}=\delta _{i}^{k}$ for the case $%
n=3$ . Since in principle the system for $n=3$ can be solved in an
elementary manner, the aim will be not to find another more convenient
method, but rather than that find a method, which can further be generalized
to higher dimensions.

In matrix notations for the unknown variables $%
g_{11},g_{22},g_{33},g_{12},g_{23}$ and $g_{13}$ the system $%
g_{ij}g^{jk}=\delta _{i}^{k}$ of nine equations (for the different indices $%
i $ and $k$) can be written as 
\begin{equation}
\widetilde{A}X=E\text{ \ \ \ ,}  \tag{3.1}
\end{equation}%
where $X^{T}$ further will denote the transponed vector \ 
\begin{equation}
X^{T}\equiv (g_{11},g_{22},g_{33},g_{12},g_{23},g_{13})\text{ \ .}  \tag{3.2}
\end{equation}

and $E^{T}$ - the transponed $9-$ vector: 
\begin{equation}
E^{T}=(1,0,0,0,1,1,0,0,1)\text{ \ \ \ \ .}  \tag{3.3}
\end{equation}%
The $6\times 9$ matrix $\widetilde{A}$ has the following interesting block
structure: 
\begin{equation}
\widetilde{A}\equiv \left( 
\begin{array}{cc}
P_{1} & Q_{1} \\ 
P_{2} & Q_{2} \\ 
P_{3} & Q_{3}%
\end{array}%
\right) \text{ \ \ \ ,}  \tag{3.4}
\end{equation}%
where ($s=1,2,3$) the matrices $P_{s}$ and $Q_{s}$ are the following: 
\begin{equation}
P_{s}\equiv \left( 
\begin{array}{ccc}
g^{s1} & g^{s2} & g^{s3} \\ 
0 & g^{s1} & 0 \\ 
0 & 0 & g^{s1}%
\end{array}%
\right) \text{ \ \ \ \ ;\ \ \ \ \ }Q_{s}\equiv \left( 
\begin{array}{ccc}
0 & 0 & 0 \\ 
g^{2s} & g^{3s} & 0 \\ 
0 & g^{2s} & g^{3s}%
\end{array}%
\right) \text{ \ \ .}  \tag{3.5}
\end{equation}%
In order to find the solution $X=\widetilde{A}^{-1}E$ of the system (3.1),
one has to find the inverse matrix $\widetilde{A}^{-1}$. For the case of
quadratic matrices with the block structure 
\begin{equation}
M\equiv \left( 
\begin{array}{cc}
A & B \\ 
C & D%
\end{array}%
\right) \text{ \ ,}  \tag{3.6}
\end{equation}%
where $A,B,C$ and $D$ are $n\times n,q\times n,n\times q$ and $q\times q$
matrices correspondingly, the so called Frobenius formulae [2] for finding
the inverse matrix $M^{-1}$ is valid 
\begin{equation}
M^{-1}=\left( 
\begin{array}{cc}
A^{-1}+A^{-1}BH^{-1}CA^{-1} & -A^{-1}BH^{-1} \\ 
-H^{-1}CA^{-1} & H^{-1}%
\end{array}%
\right) \text{ \ \ \ ,}  \tag{3.7}
\end{equation}%
where $H$ is the matrix: 
\begin{equation}
H\equiv D-CA^{-1}B\text{ \ .}  \tag{3.8}
\end{equation}%
In the present case, the Frobenius formulae cannot be applied to the block
matrix (3.4), since it is not a quadratic one. However, if $X$ is a solution
of the system (3.1), then it is a solution also of the equation $\widetilde{A%
}^{T}\widetilde{A}X=E\widetilde{A}^{T}$, where the $6\times 9$ matrix $%
\widetilde{A}$, multiplied to the left with its transponed one, gives
already the quadratic $9\times 9$ matrix $\widetilde{A}^{T}\widetilde{A}$.
Further it shall be demonstrated how the Frobenius inversion formulae can be
applied twice in respect to $\widetilde{A}^{T}\widetilde{A}$.

The matrix $\widetilde{A}^{T}\widetilde{A}$ can be calculated to be the
following block matrix: 
\begin{equation}
\widetilde{A}^{T}\widetilde{A}=\left( 
\begin{array}{cc}
\sum\limits_{i=1}^{3}P_{i}^{T}P_{i} & \sum\limits_{j=1}^{3}P_{j}^{T}Q_{j} \\ 
\sum\limits_{k=1}^{3}Q_{k}^{T}P_{k} & \sum\limits_{l=1}^{3}Q_{l}^{T}Q_{l}%
\end{array}%
\right) \text{ \ }  \tag{3.9}
\end{equation}%
and $P_{i},Q_{j}$ are the corresponding matrices (3.5) and their transponed
ones. The block matrices in (3.9) are found to be the following $3\times 3$
matrices, which shall further be identified with the corresponding block -
matrices $A,B,C$ and $D$ in (3.6): 
\begin{equation}
A\equiv \sum\limits_{i=1}^{3}P_{i}^{T}P_{i}=\left( 
\begin{array}{ccc}
\sum\limits_{i=1}^{3}(g^{i1})^{2} & \sum\limits_{i=1}^{3}g^{i1}g^{i2} & 
\sum\limits_{i=1}^{3}g^{i1}g^{i3} \\ 
\sum\limits_{i=1}^{3}g^{i1}g^{i2} & \sum\limits_{i=1}^{3}\left[
(g^{i1})^{2}+(g^{i2})^{2}\right] & \sum\limits_{i=1}^{3}g^{i2}g^{i3} \\ 
\sum\limits_{i=1}^{3}g^{i1}g^{i3} & \sum\limits_{i=1}^{3}g^{i2}g^{i3} & 
\sum\limits_{i=1}^{3}\left[ (g^{i1})^{2}+(g^{i3})^{2}\right]%
\end{array}%
\right) \text{ \ \ \ ,}  \tag{3.10}
\end{equation}%
\begin{equation}
B\equiv
\sum\limits_{j=1}^{3}P_{j}^{T}Q_{j}=\sum\limits_{j=1}^{3}g^{j1}\left( 
\begin{array}{ccc}
0 & 0 & 0 \\ 
g^{2j} & g^{3j} & 0 \\ 
0 & g^{j2} & g^{j3}%
\end{array}%
\right) \text{ \ \ \ ,}  \tag{3.11 }
\end{equation}%
\begin{equation}
C\equiv
\sum\limits_{k=1}^{3}Q_{k}^{T}P_{k}=\sum\limits_{k=1}^{3}g^{k1}\left( 
\begin{array}{ccc}
0 & g^{k2} & 0 \\ 
0 & g^{k3} & g^{k2} \\ 
0 & 0 & g^{k3}%
\end{array}%
\right) \text{ \ \ \ \ \ ,}  \tag{3.12}
\end{equation}%
\begin{equation}
D\equiv \sum\limits_{l=1}^{3}Q_{l}^{T}Q_{l}=\left( 
\begin{array}{ccc}
\sum\limits_{l=1}^{3}(g^{2l})^{2} & \sum\limits_{l=1}^{3}g^{2l}g^{3l} & 0 \\ 
\sum\limits_{l=1}^{3}g^{2l}g^{3l} & \sum\limits_{l=1}^{3}\left[
(g^{21})^{2}+(g^{3l})^{2}\right] & \sum\limits_{l=1}^{3}g^{2l}g^{3l} \\ 
0 & \sum\limits_{l=1}^{3}g^{2l}g^{3l} & \sum\limits_{i=1}^{3}(g^{3l})^{2}%
\end{array}%
\right) \text{ \ \ .}  \tag{3.13 }
\end{equation}%
Note that the diagonal block - matrices $A$ and $D$ have non - zero
determinants (even if \ $g^{ij}=dX^{i}dX^{j}$), while the non - diagonal
block - matrices $B$ and $D$ have zero - determinants. However, in order to
apply the Frobenius formulae for inverting the matrix (3.9) it is sufficient
to have as invertible only the matrix $A$ (and of course the matrix $H$).

\section{ MODIFICATION OF THE BLOCK STRUCTURE OF THE MATRIX A- THE $n=3$
CASE AND\ THE\ GENERAL $N-$DIMENSIONAL\ CASE}

The above presented method has nevertheless the following shortcomings:

1. It deals with an rectangular $p\times q$ matrix $A$ for a system of
equations with $p=\left( 
\begin{array}{c}
n \\ 
2%
\end{array}
\right) +n$ variables and $q=n^{2}$ equations. At the same time it would
have been much better to deal with a quadratic matrix at the beginning.

2. The block - matrix $A$ contains two types of matrices $P_{s}$ and $Q_{s},$%
while it would be more convenient to have just one type of an elementary
\textquotedblright constituent\textquotedblright\ $E_{k}^{(i)}$ with a
definite structure, where the indice $i$ denotes the corresponding column in
the block matrix $A$ (i.e. the number of the column, containing block -
matrices) and the indice $k$ - the corresponding row of block - matrices.

3. The matrix $Q_{s},$ given by the formulae (3.4) has a zero determinant.

To avoid these shortcomings, let us define an extended $9-$ dimensional
vector $Y$, whose transponed one is 
\begin{equation}
Y^{T}\equiv (g_{11},g_{12},g_{13},g_{21},g_{22},g_{23},g_{31},g_{32},g_{33})%
\text{ \ \ , }  \tag{4.1}
\end{equation}

where the elements $g_{21},g_{32}$ and $g_{31}$ formally shall be considered
unknown, although they are equal to their symmetric counterparts $%
g_{12},g_{23}$ and $g_{13}.$ Then the system of equations can be written as 
\begin{equation}
MY=\overline{1}\text{ \ \ \ ,}  \tag{4.2}
\end{equation}%
where $\overline{1}^{T}$ is the transponed $9$ -dimensional vector 
\begin{equation}
\overline{1}^{T}\equiv (1,0,0,0,1,0,0,0,1)  \tag{4.3}
\end{equation}%
and the $9\times 9$ matrix $M$ \ has the following block structure 
\begin{equation}
M\equiv \left( 
\begin{array}{ccc}
E_{1}^{(1)} & E_{1}^{(2)} & E_{1}^{(3)} \\ 
E_{2}^{(1)} & E_{2}^{(2)} & E_{2}^{(3)} \\ 
E_{3}^{(1)} & E_{3}^{(2)} & E_{3}^{(3)}%
\end{array}%
\right) \text{ \ \ \ \ . }  \tag{4.4 }
\end{equation}%
The elementary $3\times 3$ block matrices $E_{k}^{(1)}$, $E_{k}^{(2)}$ and $%
E_{k}^{(3)}$ ($k=1,2,3$ denotes the number of the row) in each column are
the following 
\begin{equation}
E_{k}^{(1)}\equiv \frac{1}{2}\left( 
\begin{array}{ccc}
2g^{k1} & g^{k2} & g^{k3} \\ 
0 & g^{k1} & 0 \\ 
0 & 0 & g^{k1}%
\end{array}%
\right) \text{ \ \ \ \ ,}  \tag{4.5}
\end{equation}%
\begin{equation}
E_{k}^{(2)}\equiv \frac{1}{2}\left( 
\begin{array}{ccc}
g^{2k} & 0 & 0 \\ 
g^{1k} & 2g^{2k} & g^{3k} \\ 
0 & 0 & g^{2k}%
\end{array}%
\right) \text{ \ \ \ \ ,}  \tag{4.6 }
\end{equation}%
\begin{equation}
E_{k}^{(3)}\equiv \frac{1}{2}\left( 
\begin{array}{ccc}
g^{3k} & 0 & 0 \\ 
0 & g^{3k} & 0 \\ 
g^{1k} & g^{2k} & g^{3k}%
\end{array}%
\right) \text{ \ \ \ .}  \tag{4.7}
\end{equation}%
From (4.5 - 4.7) it is seen that depending on the indice $k$ of the
elementary matrices row and of the indice $(s)$ for the elementary matrices
column, the matrix $E_{k}^{(s)}$ for the $n-$ dimensional case can be
written as 
\begin{equation}
E_{k}^{(s)}\equiv \frac{1}{(n-1)}\left( 
\begin{array}{cccccc}
g^{ks} & 0... & .... & ... & 0 & 0 \\ 
0 & g^{ks} & 0.. & ... & 0 & 0 \\ 
.... & .... & ... & .... & .... & ... \\ 
g^{k1} & g^{k2}. & .. & (n-1)g^{ks}.. & ... & g^{kn} \\ 
.... & ... & ... & ............ & .... & .... \\ 
0 & 0 & 0 & ............. & 0 & g^{ks}%
\end{array}%
\right) \text{ \ \ \ .}  \tag{4.8}
\end{equation}%
In other words, on the main diagonal of the matrix $E_{k}^{(s)}$ the
elements $g^{ks}$ are situated, on the $s-$th row - the elements $\left( 
\begin{array}{cccccc}
g^{k1}, & g^{k2},. & .. & (n-1)g^{ks}.. & ... & ,g^{kn}%
\end{array}%
\right) $ with an element $(n-1)g^{ks}$ on the $s-$th row and $s-$th column$%
. $This block structure (with some slight modifications) shall be obtained
also for the $n-$ dimensional case, and thus the $3-$ dimensional case
really helps to make the corresponding generalization for the $n-$
dimensional case.

Another advantage of the block - matrix representation is that it gives a
possibility to apply twice the Frobenius formulae. Correspondingly, in the $%
n-$ dimensional case the Frobenius formulae will be applied $(n-1)$ times.

\section{\protect\bigskip BLOCK \ MATRIX \ STRUCTURE \ IN \ THE \ $N$%
-DIMENSIONAL \ CASE}

Following the same algorithm as in the preceeding subsections, we shall try
to find the block structure of the system of equations $g_{ij}g^{jk}=\delta
_{i}^{k}$ in the $n-$dimensional case. For $i=k$ the system can be written
as $AX_{k}=\left( 
\begin{array}{c}
0 \\ 
.. \\ 
1 \\ 
...%
\end{array}%
\right) $, where $"1"$ is on the $k-$th place, the vector $X_{k}^{T}$ is 
\begin{equation}
X_{k}^{T}\equiv (g_{k1},g_{k2},........,g_{kn})\text{ \ \ \ \ }  \tag{5.1}
\end{equation}%
and the matrix $A$ is 
\begin{equation}
A\equiv \left( 
\begin{array}{cccc}
g^{11} & g^{12} & ........... & g^{1n} \\ 
g^{21} & g^{22} & ........... & g^{2n} \\ 
..... & ..... & .......... & ..... \\ 
g^{n1} & g^{n2} & ......... & g^{nn}%
\end{array}%
\right) \text{ \ \ \ }  \tag{5.2}
\end{equation}%
The corresponding vectors $X_{1},X_{2},......,X_{k}$ represent the rows of \
the symmetric matrix $N$ 
\begin{equation}
N\equiv \left( 
\begin{array}{cccc}
g_{11} & g_{12} & ...... & g_{1n} \\ 
g_{21} & g_{22} & ..... & g_{2n} \\ 
.... & .... & .... & ... \\ 
g_{n1} & g_{n2} & .... & g_{nn}%
\end{array}%
\right) \text{ \ \ \ \ ,}  \tag{5.3 }
\end{equation}%
in which the unknown variables are in the lower triangular (half) part of
the matrix (denoted by $N^{tr}$).

Let us construct a $\frac{n(n+1)}{2}\times n^{2}$ dimensional matrix $B$,
which will multiply a $\frac{n(n+1)}{2}$ dimensional vector $Y$, formed by
joining all the consequent rows of the triangular matrix $\ N^{tr}$ 
\begin{equation}
Y\equiv
(g_{11},g_{12},.......,g_{1n},g_{22},g_{23},......,g_{2n},......,g_{n1},g_{n2},.....,g_{nn})%
\text{ \ .}  \tag{5.4}
\end{equation}%
Correspondingly the matrix $B$ will have the following block triangular
structure: 
\begin{equation}
B\equiv \left( 
\begin{array}{cccc}
B_{11} & 0 & ...... & 0 \\ 
B_{21} & B_{22} & ..... & 0 \\ 
.... & .... & .... & ... \\ 
B_{n1} & B_{n2} & .... & B_{nn}%
\end{array}%
\right) \text{ \ \ \ ,}  \tag{5.5 }
\end{equation}%
where each of the block matrices $B_{kk}$ on the diagonal is an $n\times
(n-k+1)$ (i.e. $(n-k+1)$ columns and $n$ rows) dimensional matrix, obtained
from the matrix $A$ by removing the first $(k-1)$ columns. For example, $%
B_{11}\equiv A$, but 
\begin{equation}
B_{22}\equiv \left( 
\begin{array}{cccc}
g^{12} & g^{13} & ...... & g^{1n} \\ 
g^{22} & g^{23} & ..... & g^{2n} \\ 
.... & .... & .... & ... \\ 
g^{n2} & g^{n3} & .... & g^{nn}%
\end{array}%
\right) \text{ \ \ \ \ .}  \tag{5.6}
\end{equation}%
The corresponding matrix $B_{kk}$ will be 
\begin{equation}
B_{kk}\equiv \left( 
\begin{array}{cccc}
g^{1k} & g^{1,(k+1)} & .... & g^{1,n} \\ 
g^{2k} & g^{2,(k+1)} & .... & g^{2,n} \\ 
... & ........ & .... & ..... \\ 
g^{nk} & g^{n,(k+1)} & ... & g^{nn}%
\end{array}%
\right) \text{ \ \ \ \ \ .}  \tag{5.7}
\end{equation}%
The block - matrices $B_{sk}$ ($s>k)$ are $n\times (n-k+1)$ dimensional ones
with just one nonzero column (the $s-k+1$ column) with the elements $%
g^{k1},g^{k2},.......$ 
\begin{equation}
B_{sk}\equiv \left( 
\begin{array}{ccccc}
0 & ... & g^{k1} & ... & 0 \\ 
0 & .... & g^{k2} & .... & 0 \\ 
... & .... & ... & ... & ... \\ 
0 & ... & g^{k,(n-1)} & .... & 0 \\ 
0 & ... & g^{k,n} & .... & 0%
\end{array}%
\right) \text{ \ \ .}  \tag{5.8}
\end{equation}%
Let us now have a look at the complex of neighbouring block matrices around
the main block diagonal 
\begin{equation}
K^{(k)}\equiv \left( 
\begin{array}{cc}
B_{k-1,k-1} & B_{k-1,k} \\ 
B_{k,k-1} & B_{kk}%
\end{array}%
\right) \text{ \ \ \ .}  \tag{5.9}
\end{equation}%
For illustration of the block matrix multiplication and in order to derive
some useful formulaes, let us calculate $K^{(p)T}K^{(k)}$, which will be
equal to 
\begin{equation}
\left( 
\begin{array}{cc}
B_{p-1,p-1}^{T}B_{k-1,k-1}+B_{p,p-1}^{T}B_{k,k-1} & B_{p,p-1}^{T}B_{k,k} \\ 
B_{pp}^{T}B_{k,k-1} & B_{pp}^{T}B_{kk}%
\end{array}%
\right) \text{ \ \ \ \ . }  \tag{5.10}
\end{equation}%
The corresponding terms in the above matrix are: 
\begin{equation*}
B_{p-1,p-1}^{T}B_{k-1,k-1}=
\end{equation*}%
\begin{equation*}
=\left( 
\begin{array}{cccc}
g^{1,(p-1)} & g^{2,(p-1)} & .... & g^{n,(p-1)} \\ 
g^{1,p} & g^{2,p} & .... & g^{n,p} \\ 
... & ........ & .... & ..... \\ 
g^{1,n} & g^{2,n} & ... & g^{nn}%
\end{array}%
\right) \left( 
\begin{array}{cccc}
g^{1,(k-1)} & g^{1,k} & .... & g^{1,n} \\ 
g^{2,(k-1)} & g^{2,k} & .... & g^{2,n} \\ 
... & ........ & .... & ..... \\ 
g^{n,k-1} & g^{n,k} & ... & g^{nn}%
\end{array}%
\right) =
\end{equation*}%
\begin{equation}
=\left( 
\begin{array}{cccc}
P_{11} & P_{12} & ... & P_{n-k+2} \\ 
.... & .... & ... & ..... \\ 
.. & .... & P_{sr.....} & ...... \\ 
P_{n-p+2,1} & P_{n-p+2,2} & ...... & P_{n-p+2,n-k+2}%
\end{array}%
\right) \text{ \ ,}  \tag{5.11}
\end{equation}%
where all the elements of the matrix are nonzero and the element $P_{sr}$ on
the $s-$th row and on \ the $r-$th column is equal to $P_{sr}\equiv
\sum\limits_{i=1}^{n}g^{i,p+s-2}g^{i,k+r-2}$, 
\begin{equation}
B_{p,p-1}^{T}B_{k,k-1}=\left( 
\begin{array}{cccc}
0 & 0 & .. & 0 \\ 
0 & \sum\limits_{i=1}^{n}g^{(p-1),i}g^{(k-1),i} & ... & 0 \\ 
..... & ...... & .... & ... \\ 
0 & 0 & ..... & 0%
\end{array}%
\right)  \tag{5.12}
\end{equation}

where $B_{p,p-1}^{T}$ and $B_{k,k-1}$ are $(n-p+2)\times n$ $\ $and $\
n\times (n-k+2)$ matrices and the resulting $n\times n$ matrix has only one
nonzero element $\sum\limits_{i=1}^{n}g^{(p-1),i}g^{(k-1),i}$ on the second
row and on the second column. Next let us calculate the matrix $%
B_{pp}^{T}B_{k,k-1}$, which is a product of \ the \ $(n-p+1)\times n$ matrix 
$B_{pp}^{T}$ and the $\ n\times (n-k+2)$ matrix $B_{k,k-1}$ with the only
nonzero second column: 
\begin{equation}
B_{pp}^{T}B_{k,k-1}=\left( 
\begin{array}{cccc}
0 & F_{12} & ... & 0 \\ 
0 & F_{22} & ... & 0 \\ 
.... & .. & ... & ... \\ 
0 & F_{n-p+1,2} & ... & 0%
\end{array}%
\right) \text{ \ \ .}  \tag{5.13 }
\end{equation}

\bigskip The resulting matrix $B_{pp}^{T}B_{k,k-1}$ has a dimension $%
(n-p+1)\times (n-k+2)$ with the only nonzero second column with an element
on the $r-$th row and on the $2-$nd \ column $F_{r2}\equiv
\sum\limits_{i=1}^{n}g^{i,(p+r-1)}g^{k-1,i}$.

Next let us find the matrix $B_{p,p-1}^{T}B_{k,k}$, which is a product of \
the $(n-p+2)\times n$ \ matrix $B_{p,p-1}^{T}$ and the $n\times (n-k+1)$
matrix $\ B_{k,k}:$ 
\begin{equation*}
B_{p,p-1}^{T}B_{k,k}=
\end{equation*}%
\begin{equation}
=\left( 
\begin{array}{ccccc}
0 & .. & 0 & .. & 0 \\ 
\sum\limits_{i=1}^{n}g^{(p-1),i}g^{i,k} & .. & \sum%
\limits_{i=1}^{n}g^{(p-1),i}g^{i,(k+s-1)} & .. & \sum%
\limits_{i=1}^{n}g^{(p-1),i}g^{i,n} \\ 
.. & ... & .... & ... & .. \\ 
0 & .. & 0 & .. & 0 \\ 
0 & .. & 0 & .. & 0%
\end{array}%
\right) \text{ \ \ \ ,}  \tag{5.14}
\end{equation}%
where in the last formulae we have used that the obtained matrix has $%
(n-k+1) $ columns and therefore the indice $k+s-1$ in the expression for the
element $\sum\limits_{i=1}^{n}g^{(p-1),i}g^{i,(k+s-1)}$ on the $2-$nd $\ $%
row and $s- $th column ranges from $k$ $\ $to $k+s-1=k+(n-k+1)-1=n$.

\bigskip It remains only to calculate the matrix $B_{pp}^{T}B_{kk}$, but it
is the same as \ (5.11), this time with an element 
\begin{equation}
P_{sr}\equiv \sum\limits_{i=1}^{n}g^{i,p+s-1}g^{i,k+r-1}  \tag{5.15}
\end{equation}

on the $s-$th row and on the $r-$th column.

Using the above developed techniques for matrix multiplication, let us
calculate the $\frac{n(n+1)}{2}\times \frac{n(n+1)}{2}$ matrix $B^{T}B$
(recall - $B$ is an $n^{2}\times \frac{n(n+1)}{2}$ matrix and $B^{T}$ is an $%
\frac{n(n+1)}{2}\times n^{2}$ matrix), which is the $n-$dimensional analogue
of the matrix (3.8). Taking into consideration (5.9), one has 
\begin{equation}
B^{T}B=\left( 
\begin{array}{cccc}
\sum\limits_{i=1}^{n}B_{i1}^{T}B_{i1} & \sum\limits_{i=2}^{n}B_{i1}^{T}B_{i2}
& ... & \sum\limits_{i=n}^{n}B_{i1}^{T}B_{in} \\ 
\sum\limits_{i=2}^{n}B_{i2}^{T}B_{i1} & \sum\limits_{i=2}^{n}B_{i2}^{T}B_{i2}
& ... & \sum\limits_{i=n}^{n}B_{i2}^{T}B_{ni} \\ 
.... & .. & ... & ... \\ 
\sum\limits_{i=n}^{n}B_{in}^{T}B_{i1} & .. & ... & \sum%
\limits_{i=n}^{n}B_{in}^{T}B_{in}%
\end{array}%
\right) \text{ \ \ \ \ \ \ .}  \tag{5.16 }
\end{equation}%
The above matrix contains three types of elements: \ 

1-st type. Elements below the block diagonal of the type $%
\sum\limits_{\alpha =j}^{n}B_{\alpha j}^{T}B_{\alpha k}$\ with $k<j$.
Carrying out the matrix multiplication and for the moment not taking the
summation over $\alpha ,$ we find 
\begin{equation}
B_{\alpha j}^{T}B_{\alpha k}=\left( 
\begin{array}{ccccc}
0 & 0 & .. & 0 & 0 \\ 
.. & .. & .. & .. & .. \\ 
0 & .. & \sum\limits_{i=1}^{n}g^{ji}g^{ki} & .. & 0 \\ 
0 & .. & .. & .. & .0 \\ 
0 & 0 & .. & .. & 0%
\end{array}%
\right) \text{ \ \ \ \ ,}  \tag{5.17}
\end{equation}%
where the only nonzero element $\sum\limits_{i=1}^{n}g^{ji}g^{ki}$ in the
matrix is on the $(\alpha -j+1)$ row and on the $(\alpha -k+1)$ column and
the matrices $B_{\alpha j}^{T}$ and $B_{\alpha k}$ are of dimensions $%
(n-j+1)\times n$ and $n\times (n-k+1)$ correspondingly.

Since the first term in the sum $\sum\limits_{\alpha =j}^{n}B_{\alpha
j}^{T}B_{\alpha k}$ will be $B_{\alpha \alpha }^{T}B_{\alpha k}$, let us
find it, performing the same kind of matrix multiplication as in (5.13): 
\begin{equation}
B_{\alpha \alpha }^{T}B_{\alpha k}=\left( 
\begin{array}{ccccc}
0 & 0 & \sum\limits_{i=1}^{n}g^{i\alpha }g^{ki} & 0 & 0 \\ 
.. & .. & .. & .. & .. \\ 
0 & .. & \sum\limits_{i=1}^{n}g^{i,\alpha +r-1}g^{k,i} & .. & 0 \\ 
0 & .. & .. & .. & .0 \\ 
0 & 0 & \sum\limits_{i=1}^{n}g^{i,n}g^{k,i} & .. & 0%
\end{array}%
\right) \text{ \ \ \ \ ,}  \tag{5.18}
\end{equation}%
where the only nonzero column is the $(\alpha -k+1)$ one and the element in
this column and on the $r-$th row is $\sum\limits_{i=1}^{n}g^{i,\alpha
+r-1}g^{k,i}.$ The matrices $B_{\alpha \alpha }^{T}$ and $B_{\alpha k}$ are
of dimensions $(n-\alpha +1)\times n$ and $\ n\times (n-k+1)$
correspondingly and the resulting matrix $B_{\alpha \alpha }^{T}B_{\alpha k}$
is $(n-\alpha +1)\times (n-k+1)$ dimensional.

2-nd type. Elements above the block diagonal of the type $%
\sum\limits_{\alpha =k}^{n}B_{\alpha j}^{T}B_{\alpha k}$\ with $k>j$\ 
\textbf{[}The summation indice $\alpha $ takes at first the value of that
indice ($k$ or $j)$, which is greater than the other\textbf{]. }The first
term in the above sum is $B_{\alpha j}^{T}B_{\alpha \alpha }$, which can \
be found to be 
\begin{equation*}
B_{\alpha j}^{T}B_{\alpha \alpha }=
\end{equation*}%
\begin{equation}
=\left( 
\begin{array}{ccccc}
0 & 0 & .. & 0 & 0 \\ 
.. & .. & .. & .. & .. \\ 
\sum\limits_{i=1}^{n}g^{j,i}g^{1,(\alpha +i-1)} & ... & \sum%
\limits_{i=1}^{n}g^{j,i}g^{p,(\alpha +i-1)} & .. & \sum%
\limits_{i=1}^{n}g^{j,i}g^{n,(\alpha +i-1)} \\ 
.. & .. & ... & .. & .. \\ 
0 & 0 & ... & 0 & 0%
\end{array}%
\right) \text{ \ \ \ .}  \tag{5.19 }
\end{equation}%
The above $(n-j+1)\times (n-\alpha +1)$ matrix $B_{\alpha j}^{T}B_{\alpha
\alpha }$ contains a nonzero $(\alpha -j+1)$ row with an element in the $p-$%
th column, equal to 
\begin{equation}
K_{\alpha -j+1,p}=\sum\limits_{i=1}^{n}g^{j,i}g^{p,(\alpha +i-1)}\text{ \ \
\ \ .}  \tag{5.20}
\end{equation}

3-rd type. Elements situated on the block diagonal of the type\textbf{\ } 
\begin{equation}
\sum\limits_{\alpha =k}^{n}B_{\alpha k}^{T}B_{\alpha k}=B_{\alpha \alpha
}^{T}B_{\alpha \alpha }+\sum\limits_{\alpha =k+1}^{n}B_{\alpha
k}^{T}B_{\alpha k}\text{ \ .}  \tag{5.21 }
\end{equation}%
Similarly to the calculation of (5.15), the first term in (5.21) ($B_{\alpha
\alpha }^{T}$ is an $(n-\alpha +1)\times n$ \ matrix ; $B_{\alpha \alpha }$
is an $n\times (n-\alpha +1)$ matrix) is found to be the following $%
(n-\alpha +1)\times (n-\alpha +1)$ matrix (also, $\alpha =k)$ : 
\begin{equation}
B_{\alpha \alpha }^{T}B_{\alpha \alpha }=\left( 
\begin{array}{cccc}
N_{11} & N_{12} & ... & N_{1,(n-\alpha +1)} \\ 
N_{21} & N_{22} & ... & N_{2,(n-\alpha +1)} \\ 
... & ... & ... & .. \\ 
N_{n-\alpha +1,1} & ... & ... & N_{n-\alpha +1,n-\alpha +1}%
\end{array}%
\right) \text{ \ \ ,}  \tag{5.22}
\end{equation}%
where 
\begin{equation}
N_{pq}=\sum\limits_{i=1}^{n}g^{i,\alpha +p-1}g^{i,\alpha +q-1}\text{ \ \ .} 
\tag{5.23 }
\end{equation}%
The second term in (5.21) is in fact the $(n-k+1)\times (n-k+1)$ matrix
(5.17) for $j=k.$ The summation over $\alpha $ from $\alpha =k+1$ to $n$
will give a diagonal $(n-k+1)\times (n-k+1)$ matrix with an element $%
G_{\alpha -k+1,\alpha -k+1}=\sum\limits_{i=1}^{n}(g^{k,i})^{2}$ on the $%
(\alpha -k+1)$ row and on the $(\alpha -k+1)$ column, which will be situated
on the main (block\textbf{)} diagonal from $\alpha =k+1$ to $n$. Since for $%
\alpha =k+1$ we have $\alpha -k+1=k+1-k+1=2$ and for $\alpha =n$ \ we have
also $\alpha -k+1=n-k+1$, this means that the matrix $\sum\limits_{\alpha
=k+1}^{n}B_{\alpha k}^{T}B_{\alpha k}$ will have the following structure:

\begin{equation}
\sum\limits_{\alpha =k+1}^{n}B_{\alpha k}^{T}B_{\alpha k}=\left( 
\begin{array}{ccccc}
0 & 0 & .. & 0 & 0 \\ 
0 & \sum\limits_{i=1}^{n}(g^{k,i})^{2} & .. & 0 & 0 \\ 
0 & 0 & \sum\limits_{i=1}^{n}(g^{k,i})^{2} & .. & 0 \\ 
... & ... & .... & .... & ... \\ 
0 & 0 & .... & 0 & \sum\limits_{i=1}^{n}(g^{k,i})^{2}%
\end{array}%
\right) \text{ \ \ \ \ .}  \tag{5.24 }
\end{equation}%
Therefore, summing up the matrices (5.22) and (5.25), one obtains the
general structure of the matrices (5.21) $\sum\limits_{\alpha
=k}^{n}B_{\alpha k}^{T}B_{\alpha k}$\ on the block diagonal: 
\begin{equation}
\sum\limits_{\alpha =k}^{n}B_{\alpha k}^{T}B_{\alpha k}=\left( 
\begin{array}{cccc}
N_{11} & N_{12} & ... & N_{1,(n-\alpha +1)} \\ 
N_{21} & N_{22}+G_{22} & ... & N_{2,(n-\alpha +1)} \\ 
... & ... & ... & .. \\ 
N_{n-\alpha +1,1} & ... & ... & N_{n-\alpha +1,n-\alpha +1}+G_{22}%
\end{array}%
\right) \text{ \ \ ,}  \tag{5.25 }
\end{equation}%
where $N_{pq}$ and $G_{22}$ are given by expressions (5.23) and $%
G_{2,2}=\sum\limits_{i=1}^{n}(g^{k,i})^{2}$ respectively. Consequently (for $%
p\geq 2$) 
\begin{equation}
N_{pp}+G_{22}=\sum\limits_{i=1}^{n}\left[
g^{i,(k+p-1)}g^{i,(k+p-1)}+(g^{k,i})^{2}\right] \text{ \ \ \ \ \ \ .} 
\tag{5.26}
\end{equation}

Let us find now the structure of the off - diagonal block matrices, situated
below the diagonal in the block matrix (5.16). Each such an element can be
decomposed as 
\begin{equation}
\sum\limits_{\alpha =j(j>k)}B_{\alpha j}^{T}B_{\alpha k}=B_{\alpha \alpha
}^{T}B_{\alpha k}+\sum\limits_{\alpha =j+1}^{n}B_{\alpha j}^{T}B_{\alpha k}%
\text{ \ \ \ .}  \tag{5.27}
\end{equation}%
The first term in (5.27) is the already known $(n-\alpha +1)\times (n-\alpha
+1)$ matrix (5.18) with the only nonzero $(j-k+1)$ column. The second term
is the sum from $\alpha =j+1$ to $\alpha =n$ of the $(n-\alpha +1)\times
(n-k+1)$ matrices $B_{\alpha j}^{T}B_{\alpha k}$, already calculated in
(5.17) and having the only nonzero element $\sum%
\limits_{i=1}^{n}g^{ji}g^{ki} $ on the $(\alpha -j+1)$ row and on the $%
(\alpha -k+1)$ column. The summation over $\alpha $ from $\alpha =j+1$ to $%
\alpha =n$ means that in the sum $\sum\limits_{\alpha =j+1}^{n}B_{\alpha
j}^{T}B_{\alpha k}$ the element $\sum\limits_{i=1}^{n}g^{ji}g^{ki}$ will
appear beginning from the row $\alpha -j+1=2$ (for $\alpha =j+1$) up to the
row $\alpha -j+1=n-j+1$ (for $\alpha =n$), which in fact is the last row%
\textbf{. }Correspondingly, the same element will appear beginning from the
column $\alpha -k+1=j-k+2$ (for $\alpha =j+1)$ and ending at the column $%
\alpha -k+1=n-k+1$ (at $\alpha =n$), which is the last column. In other
words, the summation of the matrices $B_{\alpha j}^{T}B_{\alpha k},$%
containing one element, effectively results in a matrix, filled up from the
second row to the end and from the $(j-k+2)$ column to the end: 
\begin{equation*}
\sum\limits_{\alpha =j+1}^{n}B_{\alpha j}^{T}B_{\alpha
k}=\sum\limits_{\alpha =j+1}^{n}\left( 
\begin{array}{ccccc}
0 & .. & 0 & .. & 0 \\ 
... & .. & .... & ... & ... \\ 
0 & .... & \sum\limits_{i=1}^{n}g^{ji}g^{ki} & .... & 0 \\ 
... & ... & ... & ... & ... \\ 
0 & .. & 0 & ... & 0%
\end{array}%
\right) =
\end{equation*}%
\begin{equation}
=\left( 
\begin{array}{ccccc}
0 & ... & \text{column }(j-k+2)... & 0 & 0 \\ 
0 & ... & \sum\limits_{i=1}^{n}g^{ji}g^{ki} & .. & \sum%
\limits_{i=1}^{n}g^{ji}g^{ki} \\ 
.. & .. & .. & .. & .. \\ 
0 & .. & \sum\limits_{i=1}^{n}g^{ji}g^{ki} & .. & \sum%
\limits_{i=1}^{n}g^{ji}g^{ki} \\ 
0 & .. & \sum\limits_{i=1}^{n}g^{ji}g^{ki} & .. & \sum%
\limits_{i=1}^{n}g^{ji}g^{ki}%
\end{array}%
\right) \text{ \ \ .}  \tag{5.28}
\end{equation}%
Now recall that the matrix $B_{\alpha \alpha }^{T}B_{\alpha k}$ (5.18) had a
nonzero $(j-k+1)$ column, so therefore the structure of the whole matrix $%
\sum\limits_{\alpha =j}^{n}B_{\alpha j}^{T}B_{\alpha k}$ ($j>k$) below the
diagonal is similar to (5.28), but with the additional $(j-k+1)$ column: 
\begin{equation}
\left( 
\begin{array}{cccccc}
0 & 0.......... & \sum\limits_{i=1}^{n}g^{i,\alpha }g^{k,i} & 0 & .. & 0 \\ 
0 & 0...... & \sum\limits_{i=1}^{n}g^{i,\alpha +1}g^{k,i} & 
\sum\limits_{i=1}^{n}g^{ji}g^{ki} & ... & \sum\limits_{i=1}^{n}g^{ji}g^{ki}
\\ 
... & ... & ..... & ... & ... & ... \\ 
0 & 0........ & \sum\limits_{i=1}^{n}g^{i,\alpha +r-1}g^{k,i} & 
\sum\limits_{i=1}^{n}g^{ji}g^{ki} & .. & \sum\limits_{i=1}^{n}g^{ji}g^{ki}
\\ 
.. & ..... & .. & .. & ... & .. \\ 
0 & 0..... & \sum\limits_{i=1}^{n}g^{i,n}g^{k,i} & \sum%
\limits_{i=1}^{n}g^{ji}g^{ki} & .. & \sum\limits_{i=1}^{n}g^{ji}g^{ki}%
\end{array}%
\right) \text{ \ \ .}  \tag{5.29}
\end{equation}

\ \ \ In a completely analogous way the elements above the block diagonal in
(5.16) can be found. These elements $\sum\limits_{\alpha =k}^{n}B_{\alpha
j}^{T}B_{\alpha k}$ ($k>j$) can be decomposed as 
\begin{equation}
\sum\limits_{\alpha =k(k>j)}B_{\alpha j}^{T}B_{\alpha
k}=B_{kj}^{T}B_{kk}+\sum\limits_{\alpha =k+1}^{n}B_{\alpha j}^{T}B_{\alpha k}%
\text{ \ \ \ .}  \tag{5.30}
\end{equation}%
This formulae is similar to (5.27) for the below - diagonal elements, but
here in (5.30) we have the matrix $B_{kj}^{T}B_{kk}$ instead of the matrix $%
B_{\alpha \alpha }^{T}B_{\alpha k}$ and the summation over the indice $%
\alpha $ in the second term is from $\alpha =k+1$ to $\alpha =n$ instead of $%
\alpha =j+1$ to $\alpha =n$ in (5.27). The first term in (5.30) $%
B_{kj}^{T}B_{kk}$ is the $(n-j+1)\times (n-k+1)$ matrix (5.19) \ (with $%
\alpha =k$) with the only nonzero $(k-j+1)$ row with the elements 
\begin{equation}
\sum\limits_{i=1}^{n}g^{ji}g^{1,(k+i-1)}\text{ \ \ ; ....}%
\sum\limits_{i=1}^{n}g^{ji}g^{p,(k+i-1)}\text{;.....}\sum%
\limits_{i=1}^{n}g^{ji}g^{n,(k+i-1)}\text{\ \ \ }  \tag{5.31}
\end{equation}

The second term in (5.30) is again the sum of the matrices (5.17). The only
nonzero element $\sum\limits_{i=1}^{n}g^{ji}g^{ki}$ will now appear in the
final sum from the $\alpha -j+1=k-j+2$ $\ $row \ ($\alpha =k+1$) until the $%
k-j+1=n-j+1$ ($\alpha =n$) row, which is the last one. Also, the same
element will fill up the columns from $\alpha -k+1=2$ $\ $($\alpha =k+1$) to
the column $\alpha -k+1=n-k+1$ ($\alpha =n$), which is also the last one.
Therefore, summing up the two terms in (5.30), one obtains the following
matrix for the above - diagonal block terms\textbf{, }in which the $(k-j+1)$
row is filled up with the elements $\sum\limits_{i=1}^{n}g^{j,i}g^{p,(\alpha
+i-1)}$ and from the next $(k-j+2)$ row to \ the end and from the second
column to the end column the matrix is filled up with the other element $%
\sum\limits_{i=1}^{n}g^{ji}g^{ki}:$ 
\begin{equation}
\left( 
\begin{array}{cccccc}
0 & 0.......... & 0.. & 0.. & .. & 0 \\ 
.. & ...... & ... & ... & ... & .... \\ 
\sum\limits_{i=1}^{n}g^{j,i}g^{1,(\alpha +i-1)} & \sum%
\limits_{i=1}^{n}g^{j,i}g^{2,(\alpha +i-1)} & ........\sum%
\limits_{i=1}^{n}g^{j,i}g^{p,(\alpha +i-1)} & ... & ... & 
\sum\limits_{i=1}^{n}g^{j,i}g^{n,(\alpha +i-1)} \\ 
0 & \sum\limits_{i=1}^{n}g^{ji}g^{ki} & ..... & \sum%
\limits_{i=1}^{n}g^{ji}g^{ki} & .. & \sum\limits_{i=1}^{n}g^{ji}g^{ki} \\ 
.. & ..... & .. & .. & ... & .. \\ 
0 & \sum\limits_{i=1}^{n}g^{ji}g^{ki} & \sum\limits_{i=1}^{n}g^{i,n}g^{k,i}
& \sum\limits_{i=1}^{n}g^{ji}g^{ki} & .. & \sum\limits_{i=1}^{n}g^{ji}g^{ki}%
\end{array}%
\right) \text{ \ \ .}  \tag{5.32 }
\end{equation}

\bigskip Let us now summarize the obtained results for the $n-$dimensional
case. The (predetermined) system of equations $g_{ij}g^{jk}=\delta _{i}^{k}$
was represented as $BY=\overline{1}$ , where \ $\overline{1}$ is an $n^{2}$
dimensional vector, whose transponed one is defined as $\overline{1}^{T}=(%
\overline{1}^{T1},\overline{1}^{T2},.............,\overline{1}^{Tn})$ and
the corresponding transponed $n-$ dimensional vectors $\overline{1}^{T1},%
\overline{1}^{T2},.............,\overline{1}^{Tn}$ are defined as follows: $%
\overline{1}^{T1}=(1,0,........,0),\overline{1}^{T2}=(0,1,0,........0)$ and
the $k-$th transponed vector $\overline{1}^{Tk}=(0,0,.....,1,0,...0)$
contains the number $1$ on the $k-$th place. The $n^{2}\times \frac{n(n+1)}{2%
}$ matrix $B$ in terms of the elementary block matrices $B_{ij}$ and the $%
\frac{n(n+1)}{2}$ dimensional vector $Y$ were defined by formulaes (5.5 -
5.8) and (5.4) respectively. In order to solve the system, we multiplied it
to the left with the transponed matrix $B^{T}$ and thus the solution for the
vector $Y$ in matrix notations can be found as $Y=(B^{T}B)^{-1}B^{T}%
\overline{1}$, where the expression for $B^{T}\overline{1}$ can easily be
found to be 
\begin{equation}
B^{T}\overline{1}=\left( 
\begin{array}{cccc}
B_{11}^{T} & B_{21}^{T} & ...... & B_{n1}^{T} \\ 
0 & B_{22}^{T} & .... & B_{n2}^{T} \\ 
... & ... & ... & .. \\ 
0 & 0 & .. & B_{nn}^{T}%
\end{array}%
\right) \left( 
\begin{array}{c}
\overline{1}^{1} \\ 
\overline{1}^{2} \\ 
... \\ 
\overline{1}^{m}%
\end{array}%
\right) =\left( 
\begin{array}{c}
\sum\limits_{i=1}^{n}B_{i1}^{T}\overline{1}^{i} \\ 
\sum\limits_{i=1}^{n}B_{i2}^{T}\overline{1}^{i} \\ 
... \\ 
\sum\limits_{i=1}^{n}B_{im}^{T}\overline{1}^{i}%
\end{array}%
\right) \text{ \ \ \ .}  \tag{5.33}
\end{equation}%
The expressions for the elements of the vector in the R. H. S. can also be
found, but this will not be performed here and will be left for the
interested reader. Let us remind again \ that each \textquotedblright
element\textquotedblright\ of this vector is in itself an $n-$dimensional
vector and $m=\frac{n(n+1)}{2}$ gives the number of the \textquotedblright
block\textquotedblright\ elements of this vector - column. The number $m$
should be an integer number, but this requirement can be fulfilled and this
will be commented in the next subsection.

The main result of this subsection are contained in the expressions (5.25),
(5.29) and (5.32) for the elements of the matrix $B^{T}B$\ on the block
diagonal, below the block diagonal and above the block diagonal
respectively. In such a way the detailed structure of the matrtix $B^{T}B$\
is known in terms of the elementary constituent block matrices $B_{ij}$. \
The structure of the matrix $B^{T}B$ is important for the following
reasons:\ 

1. If one chooses the contravariant metric tensor in the form of a
factorized product $\widetilde{g}^{ij}=dX^{i}dX^{j}$, then one can answer
the question whether it is possible the rows of the matrix $B^{T}B$ to have
some common multiplier. But from the above expressions and since in each
\textquotedblright cell\textquotedblright\ of the elementary block matrices
one has a summation of the kind $\sum\limits_{i=1}^{n}g^{i,n}g^{k,i}$, it is
evident that such a common multiplier does not exist and therefore the rank
of the matrix $B^{T}B$ cannot be lowered due to the above made choice of $%
g^{ij}$.

2. The quadratic structure of \ the \ matrix $B^{T}B$ gives an opportunity \
to apply the Frobenius formulae for finding the inverse matrix $%
(B^{T}B)^{-1},$as previously discussed.

\section{\protect\bigskip BLOCK MATRIX REPRESENTATION OF THE HOMOGENEOUS
SYSTEM \ OF \ EQUATIONS WITH \ A \ ZERO \ R. H. S. \ }

Earlier it was shown that for the case $n=3$ the sub - system of equations $%
g_{ij}g^{jk}=0$ with a zero R. H. S. has a determinant of coefficient
(functions), equal to zero. The question which naturally arises is whether
this property is valid only for the $n=3$\ case and is it valid for the $n-$%
\ dimensional case?

Below it shall be proved that this can be done for the general case.\
Namely, it shall be established that from the $n^{2}-n$ equations with a
zero R. H. S. there may be chosen a sub-system of $\left( 
\begin{array}{c}
n \\ 
2%
\end{array}%
\right) +n=\frac{n(n+1)}{2}$ equations with a determinant of coefficients,
equal to zero. It shall be stressed that the proof will be that such a
system exists (i.e. can be chosen) and not that all other choices of the
subsystem of equations will also satisfy this requirement. In fact,
investigating under what other choices of the sub - system of equations with
a zero R. H. S. this property will \ be preserved, represents an interesting
problem for further research.

For the purpose, let us again use the block matrix representation (5.5).
Then, in order to obtain the system of equations with a zero R. H. S. , one
has to remove the first row (for $i=k=1$) from the system $%
g^{nk}g_{ki}=\delta _{i}^{n}$ of the first $n$ equations (i.e. for $n=i$),
then the second row from the second system of $n$ equations and so on, one
has to remove the $k-$th row from the $k-$th system of $n$ equations.
Further, in order to receive again a block matrix structure with
(elementary) submatrices with $n$ rows, one should add the first row of the
second system of $n$ equations as the last row of the first system of
equations. In effect, since the first two rows of the second system have
been removed, one should add the first two rows of the third system of $n$
equations as the last two rows of the second system. Therefore, since also
the third row in the third system ( for $i=k=3$) has been removed, it has a
total of three rows removed, and subsequently three last rows have to be
added from the fourth system. Continuing in the same manner, from the $k-$th
matrix $B_{kk}$ on the block diagonal we have the first $k-$ rows from $%
B_{kk}$ removed and also $k$ - last rows added, which should be taken from
the below - diagonal matrix $B_{k+1,k}$. Since according to (5.8) the $%
n\times (n-k+1)$ matrix $B_{sk\text{ }}$has a nonzero $(s-k+1)$ column, the
matrix $B_{k+1,k}$ will have a nonzero $k+1-k+1=2$ column. Therefore the new
transformed in this way $n\times (n-k+1)$ matrix, which will be denoted as $%
\widetilde{B}_{kk}$ , will have the following structure:\ 
\begin{equation}
\widetilde{B}_{kk}\equiv \left( 
\begin{array}{ccccccc}
g^{k+1,k} & g^{k+1,k+1} & .. & .. & ... & .. & g^{k+1,n} \\ 
g^{k+2,k} & g^{k+2,k+1} & .. & .. & . & . & g^{k+2,n} \\ 
.. & .. & .. & .. & . & .. & ... \\ 
g^{nk} & g^{n,(k+1)} & .. & .. & .. & .. & g^{nn} \\ 
0 & g^{k1} & .. & 0 & .. & 0 & 0 \\ 
.. &  & .. & .. & .. & .. & .. \\ 
0 & g^{kk} & .. & 0 & 0 & 0 & 0%
\end{array}%
\right)  \tag{6.1}
\end{equation}%
In the same way, the below - diagonal transformed matrix $\widetilde{B}%
_{k+1,k}$, obtained from $B_{k+1,k}$ by removing its first $k$ and adding $k$
rows from $B_{k+2,k}$ , will be of the following kind:

\begin{equation}
\widetilde{B}_{k+1,k}\equiv \left( 
\begin{array}{cccccc}
0 & g^{k,(k+1)} & 0 & .. & 0 & 0 \\ 
.. & .. & .. & .. & .. & .. \\ 
0 & g^{k,n} & 0 & .. & 0 & 0 \\ 
0 & 0 & g^{k,1} & .. & 0 & 0 \\ 
.. & .. & .. & ... & .. & .. \\ 
0 & 0 & g^{k,k} & .. & 0 & 0%
\end{array}%
\right) \text{ \ \ \ \ .} \ 
\end{equation}
Since the block matrix (5.5) has a \textbf{t}riangular structure, for our
further purposes only the structure of the block - diagonal matrices $%
\widetilde{B}_{kk}$ will be relevent.

Next, our \ goal will be to divide the the block - matrix (5.5) into
elemetary block matrices with an equal number ($n$) of rows and columns. Let
us remind once again that the block - matrix (5.5) contained elementary
block - matrices with an unequal number of columns - $n,(n-1),(n-2)......$%
correspondingly. For the purpose, we shall take one (left) column from the
block matrix with $(n-2)$ columns and transfer it to the left to the block
matrice with $(n-1)$ columns. As a result, the block matrices on the second
block matrix column (B. M. C.) will already contain $n$ columns. Since in
the block matrix column one column has been transfered, one has to add $3$
columns from the $(n-3)-$rd block matrix column to the $(n-2)-$nd \ block
matrix column in order to obtain again a block matrix column, consisting of
elementary $n\times n$ matrices. Continuing in the same manner with the $%
(n-3)-$rd \ B. M. C., one has to add to its right end $6$ columns from the $%
(n-4)-$th B. M. C. Now let us write down the numbers of the corresponding
block columns and below with a $(-)$ sign the number of columns, transferred
to the (neighbouring) B. M.\ C. ; with a $(+)$ sign the number of columns,
joined to the B. M.C. (to the right side) from the neighbouring (right) B.
M. C. 
\begin{equation*}
(n-1)\text{ \ \ \ \ \ ; \ \ \ \ \ }(n-2)\text{ \ \ \ \ ; \ \ \ \ \ \ \ \ \ \
\ \ \ \ }(n-3)\text{ ; \ \ \ \ \ \ \ \ \ \ \ \ \ \ \ \ \ }(n-4)\text{ \ \ \
; \ \ \ \ \ \ \ \ \ \ \ \ \ \ \ \ }(n-5)
\end{equation*}
\begin{equation*}
+1\text{; \ \ \ \ \ \ \ \ \ \ \ \ \ }-1\text{ \ \ \ \ \ \ }+3\text{ \ \ \ \
\ \ \ ; \ \ }-3\text{ \ \ \ \ \ \ }+6\text{ \ \ \ \ ; \ }-6\text{ \ \ \ \ \
\ \ }+10\text{ \ \ \ ; \ \ \ \ }-10\text{ \ \ \ \ \ \ \ \ }+15\text{ \ \ \ \
\ \ \ \ \ \ .}
\end{equation*}
Now let us look at the numbers with a minus sign, which form the following
number sequence (with the corresponding number in the sequence denoted):

\begin{equation*}
1,3,6,10,15,21......
\end{equation*}%
\begin{equation}
\text{\ \ \ \ \ \ \ \ \ \ \ \ \ \ \ \ \ \ \ \ \ \ \ \ \ \ \ \ }\ \ \text{\ \
\ \ \ \ }1,2,3,4,5,6,......\text{\ \ \ \ \ \ \ \ \ \ \ \ \ \ \ \ \ \ \ \ \ \
\ \ \ \ \ \ \ \ \ \ \ \ \ \ \ }  \tag{6.2}
\end{equation}

It is trivial to note that each number in the sequence (upper row) is in
fact a sum of the corresponding numbers in the lower row up to that number.
For example, the number $21$ in the sequence (upper row) can be represented
as a sum of the numbers in the sequence (lower row): $21=1+2+3+4+5+6$. The
same with the number $10$ $\Longrightarrow 10=1+2$\bigskip $+3+4$.
Therefore, to the sequence number $k$ in the low row will correspond the
number $\frac{k(k+1)}{2}$ in the upper row, which is the sum of the first $k$
numbers in the low row. Since the number $k$ in the lower (6.2) corresponds
to the $(n-k)-$th block column, the corresponding number will be $\frac{%
k(k-1)}{2}$, and it will correspond to the number of columns, which have to
be removed from the diagonal block matrix $B_{k+1,(k+1)}.$ At the same time,
to the right end one should add $k+\frac{k(k-1)}{2}$ $=\frac{k(k+1)}{2}$
columns. This number is exactly equal to the number of left columns, removed
from the (right) neighbouring matrix $B_{(k+1),(k+2)}.$ This can serve also
as a consistency check that the performed calculation is consistent and
correct.

From the matrix (6.1) for $\widetilde{B}_{k+1,(k+1)}$ we have to delete the
first $\overline{s}+1$ left columns, the first (upper) elements of which
begin with the elements $g^{(k+2),(k+1)};$ $%
g^{(k+2),(k+2)}.............,g^{(k+2),(\overline{s}+k)}$, where $\overline{s}%
=$ $\frac{k(k-1)}{2}$. Now let us denote by $\overline{B}_{(k+1),(k+1)\text{ 
}}$ the matrix $\widetilde{B}_{k+1,(k+1)}$ with $\frac{k(k+1)}{2}$ left
columns removed and $\frac{k(k+1)}{2}$ right columns added. The upper
elements of the (left) remaining columns will be $g^{(k+2),(\overline{s}%
+k+1)},g^{(k+2),(\overline{s}+k+2)},........,g^{(k+2),n},$ where 
\begin{equation}
\overline{s}+k+1=\frac{k(k-1)}{2}+k+1=\frac{k(k+1)}{2}+1\text{ }  \tag{6.3 }
\end{equation}%
and the matrix $\overline{B}_{(k+1),(k+1)\text{ }}$ will contain $n-\frac{%
k(k+1)}{2}$ left nonzero columns. Therefore the $n\times n$ matrix $%
\overline{B}_{(k+1),(k+1)\text{ }}$ will have the following structure: 
\begin{equation}
\overline{B}_{(k+1),(k+1)\text{ }}\equiv \left( 
\begin{array}{cc}
L_{1} & L_{2} \\ 
L_{3} & L_{4}%
\end{array}%
\right) \text{ \ \ \ ,}  \tag{6.4 }
\end{equation}%
where the $(n-k+1)\times \left[ n-\frac{k(k+1)}{2}\right] $ matrix $L_{1}$
is 
\begin{equation}
L_{1}\equiv \left( 
\begin{array}{cccc}
g^{(k+2),(\frac{k(k+1)}{2}+1)} & ... & ... & g^{(k+2),n} \\ 
g^{(k+3),(\frac{k(k+1)}{2}+1)} & ... & .... & g^{(k+3),n} \\ 
.... & ... & ... & ... \\ 
g^{n,\frac{k(k+1)}{2}+1} & ... & ... & g^{n,n}%
\end{array}%
\right)  \tag{6.5}
\end{equation}%
and $L_{2},L_{3}$ and $L_{4}$ are zero matrices of dimensions $(n-k+1)\times %
\left[ \frac{k(k-1)}{2}\right] ,(k+1)\times \left[ n-\frac{k(k+1)}{2}\right] 
$ and $(k+1)\times \left[ \frac{k(k+1)}{2}\right] $ correspondingly. Note
also that the block matrices $L_{1}$ and $L_{3}$ contain $\left[ n-\frac{%
k(k+1)}{2}\right] $ nonzero columns, and the remaining $\left[ \frac{k(k+1)}{%
2}\right] $ columns of the matrices $L_{2}$ and $L_{4}$ are exactly equal to
the number of zero columns, added to the right side of the matrix $\overline{%
B}_{(k+1),(k+1)\text{ }}$ from the neighbouring matrix $\overline{B}%
_{(k+1),(k+2)\text{ }}.$ Since this result depends on the initial structure
of the matrix $B_{(k+1),(k+1)\text{ }}$ and on the expression (6.3) (which
are both independent on the number of removed right columns), this also
confirms the consistency of the calculation.

Note that the block structure \ of the matrix (6.4) has been revealed on the
base of the assumption that the elements $g^{(k+1),1},...g^{(k+1),(k+1)}$ in
the last $(k+1)$ rows and the second column in the matrix (6.1) will be
among the first removed to the left (and outside the matrix) columns.
However, for $\overline{s}=0$ (i.e. $k=1$) the elements in the second column
of the last two rows of the matrix $\overline{B}_{2,2}$ will contain the
elements $g^{21},g^{22}.$ Therefore, after removing the first column in the
matrix (6.1) $\widetilde{B}_{(k+1),(k+1)\text{ }}($for $k=1$) and adding to
the right one zero column, the obtained structure of the $n\times n$ matrix $%
\overline{B}_{2,2}$ will be the following: 
\begin{equation}
\overline{B}_{2,2}=\left( 
\begin{array}{cccccc}
g^{33} & g^{34} & ... & g^{3n} & 0 & 0 \\ 
g^{43} & g^{44} & ... & g^{4n} & 0 & 0 \\ 
.. & ... & .. & ... & .. & .. \\ 
g^{n3} & g^{n4} & ... & g^{nn} & 0 & 0 \\ 
g^{21} & 0 & .. & 0 & 0 & 0 \\ 
g^{22} & 0 & ... & 0 & 0 & 0%
\end{array}%
\right) \text{ \ .}  \tag{6.6}
\end{equation}%
Having established the block structure of the matrix of coefficients of the $%
n-$dimensional predetermined system $g_{ij}g^{jk}=0$ with a homogeneous zero
R. H. S., it is now easy to show that $\frac{n(n+1)}{2}$ equations can be
chosen so that the determinant of coefficients will be zero. Let us take for
example the first $\frac{n(n+1)}{2}$ equations from the system of $(n^{2}-n)$
equations with a zero R. H. S. with the corresponding $\frac{n(n+1)}{2}%
\times \frac{n(n+1)}{2}$ dimensional block matrix. In the particular case
the dimension of the block matrix is determined by the number of block
matrices on the horizontal ($\frac{n(n+1)}{2}$) and on the vertical ($\frac{%
n(n+1)}{2})$. Therefore, outside this matrix will remain a matrix of $\frac{%
n(n+1)}{2}$ matrix block columns and $(n-1)-\frac{n(n+1)}{2}=\frac{(n-3)}{2}$
block rows. The last assumption presumes that the spacetime dimension number 
$n$ is an odd one, so that $(n+1)$ and $(n-3)$ are dividable by two.
Otherwise, if $n$ is an even number, one may consider $\frac{n}{2}\times 
\frac{n}{2}$ dimensional elementary block matrices $\overline{B}_{k,k}$ .
Then the full block matrix of the system will have $(n+1)$ block matrices on
the block horizontal (i.e. $(n+1)$ block columns) and $2n^{2}$ matrices on \
the block vertical (i. e. $2n^{2}$ block rows). The block matrix of the
homogeneous system of equations (with a zero R. H. S. ) will be $%
2(n-1)\times (n+1)$ \textquotedblright block\textquotedblright\ dimensional.
The chosen block matrix will be $(n+1)\times (n+1)$ \textquotedblright
block\textquotedblright\ dimensional. Outside this matrix there will remain
a block matrix of $2n-2-n-1=n-3$ block rows and $(n+1)$ block columns.

Let us now compute the determinant of the triangular matrix (5.5), from
which we take the first $\frac{(n+1)}{2}$ (or $\frac{n}{2}$) block rows.
This $\frac{(n+1)}{2}\times \frac{(n+1)}{2}$ (or $\frac{n}{2}\times \frac{n}{%
2})$ block determinant $\widetilde{S}$ will be equal to 
\begin{equation}
\widetilde{S}=\prod\limits_{i=1}^{\frac{n+1}{2}}(det\overline{B}_{ii})\text{
\ \ \ .}  \tag{6.7 }
\end{equation}%
But for $i\neq 1,2$ the expression for $det\overline{B}_{ii}$ has to be
found from formulae (6.3) for the block matrix $\overline{B}_{kk}$. Since
only one of the submatrices $L_{1}$ is different from zero, it is clear that 
$det\overline{B}_{ii}=detL_{1}.0=0$, therefore the whole expression (6.7)
equals zero.

Thus we have proved that by taking $\frac{(n+1)}{2}$ (or $\frac{n}{2})$
consequent block rows \ from the initial (quadratic and triangular) block
matrix, the obtained block matrix will have a zero determinant.

\section{DISCUSSION}

The transformation of a matrix system of algebraic equations (with the
unknown matrix $A$ assumed to be a symmetric one) into a system of linear
equations is related to some new problems of algebraic nature. The fact that
the initial matrix system is a predetermined one and solutions are not
unique should be reflected in the finally obtained matrix $B^{T}B$, but
presently it is not known how the properties of the initial matrix system
are "encoded" in this matrix.

It is interesting to mention about some possible physical applications of
the proposed mathematical algorithm. In theories with extra dimensions, the
graviton's field components $h_{MN}(x)$ in the Kaluza-Klein's expansion
under toroidal compactification along the extra-dimensional coordinate $y$
are contained the following fields [3]: 
\begin{equation}
\text{the radion \ }H^{k}=\frac{1}{k}h_{j}^{\overrightarrow{k}j}\text{ \ , }
\tag{7.1}
\end{equation}%
\begin{equation}
\text{the scalars \ }S_{ij}^{\overrightarrow{k}}=h_{ij}^{\overrightarrow{k}}-%
\frac{k}{(n-1)}(\eta _{ij}+\frac{k_{i}k_{j}}{k^{2}})H^{k}\text{ ,}  \tag{7.2}
\end{equation}%
\begin{equation}
\text{the vectors \ }V_{\mu j}^{\overrightarrow{k}}=\frac{i}{\sqrt{2}}h_{\mu
j}^{\overrightarrow{k}}\text{ \ \ ,}  \tag{7.3}
\end{equation}%
\begin{equation}
\text{the gravitons }G_{\mu \nu }^{\overrightarrow{k}}=h_{\mu \nu }^{k}+%
\frac{k}{3}(\eta _{\mu \nu }+\frac{\partial _{\mu }\partial _{\nu }}{k^{2}}%
)H^{\overrightarrow{k}}\text{ \ . }  \tag{7.4}
\end{equation}%
These fields constitute the $(4+n)\times (4+n)$ dimensional matrix of the
bulk graviton 
\begin{equation}
\left( 
\begin{array}{cc}
G_{\mu \nu }^{\overrightarrow{k}} & V_{\mu j}^{\overrightarrow{k}} \\ 
V_{\mu j}^{\overrightarrow{k}} & S_{ij}^{\overrightarrow{k}}%
\end{array}%
\right) \text{ \ \ ,}  \tag{7.5}
\end{equation}%
where $\overrightarrow{k}$ is a $n-$component vector, denoting the
Kaluza-Klein (KK) numbers along the various extra dimensions. Clearly the
matrix (7.5) is a block-structured one and the elements in it satisfy the
following constraints 
\begin{equation}
(\square +\widehat{k}^{2})G_{\mu \nu }^{\overrightarrow{k}}=0\text{ \ \ ; \ }%
\partial ^{\mu }G_{\mu \nu }^{\overrightarrow{k}}=0\text{ \ ; \ \ }G_{\mu
}^{\mu \overrightarrow{k}}=0  \tag{7.6}
\end{equation}%
\begin{equation}
\widehat{k}^{j}V_{\mu j}^{\overrightarrow{k}}=0\text{ \ \ \ , }\partial
^{\mu }V_{\mu j}^{\overrightarrow{k}}=0\text{ ,}  \tag{7.7}
\end{equation}%
\begin{equation}
k^{j}S_{ik}^{\overrightarrow{k}}=0\text{ \ ,\ }S_{k}^{\overrightarrow{k}j}=0%
\text{\ .}  \tag{7.8}
\end{equation}%
The operators $(\square +\widehat{k}^{2})$ and $\partial ^{\mu }$ may be
assumed to have some $k-$representation. Then the system (7.6-7.8) is
similar to the homogeneous part (when $i\neq k$) of the investigated system $%
g_{ij}g^{jk}=\delta _{i}^{k}$. Since the constraints are acting on a part of
the elements of the block-matrix (7.5), it will be predetermined also.

Making use of the defining equalities (7.1-7.4) for the radion, the scalars,
the vectors and the gravitons, the system of equations (7.6-7.8) can be
written with respect to the graviton's field components $h_{MN}(x\dot{)}$.
From a physical point of view, a solution in terms of these components may
be much more valuable.

The method, however, will require a modification of the approach in the
preceeding sections, since there it was assumed that the matrix of the
coefficient functions $A$ is a symmetrical one ($g^{ij}=g^{ji}$), while
finding out the graviton's components will be based on an initial matrix
system, which will be a non-symmetric one.

\section*{Acknowledgments}

The author is grateful to Dr. L. K. Alexandrov, Dr. I. Pestov, Dr. D. M.
Mladenov, St. Mishev, and especially to Prof. V. V. Nesterenko (LTPh, JINR,
Dubna)\ and Dr. O. Santillan for valuable comments, discussions and critical
remarks. \ 

\


\begin{thebibliography}{9}
\bibitem{[1]} P. Bergmann, \textquotedblright Introduction to the Theory of
Relativity\textquotedblright , 1942

\bibitem{[2]} F. R. Gantmacher 1959 \textit{Theory of Matrices} (Chelsea)

\bibitem{[3]} C. Csaki 2004 TASI Lectures on Extra Dimensions and Branes (%
\textit{Lectures at the Theoretical Advanced Study Institute, University of
Colorado, Boulder, CO June 3-28, 2002}) (\textit{Preprint} hep-ph/0404096)
\end{thebibliography}
\end{document}